\documentclass[12pt,twocolumn,preprint, a4]{iopart}
\usepackage{graphicx}
\usepackage{booktabs}
\usepackage{textcomp}
\usepackage{color}

\begin{document}

\title[Time resolved measurement of film growth during reactive HIPIMS sputtering]
	{Time resolved measurement of film growth during reactive high
       power pulsed magnetron sputtering (HIPIMS) of titanium nitride}

\author{F. Mitschker, M. Prenzel, J. Benedikt, C. Maszl and\\ A. von Keudell\footnote{\ead{Achim.vonKeudell@rub.de}}}

\address{Research Department Plasmas with Complex Interactions,
Ruhr-Universit\"at Bochum, Institute for Experimental Physics II,
D-44780 Bochum, Germany}

\date{\today}

\begin{abstract}
The growth rate during reactive high power pulsed magnetron
sputtering (HIPIMS) of titanium nitride is measured with a temporal
resolution of up to 25~$\mu$s using a rotating shutter concept.
According to that concept a 200~$\mu$m slit is rotated in front of
the substrate synchronous with the HIPIMS pulses. Thereby, the
growth flux is laterally distributed over the substrate. By
measuring the resulting deposition profile with profilometry and
with x-ray photoelectron spectroscopy, the temporal variation of
the titanium and nitrogen growth flux per pulse is deduced. The
analysis reveals that film growth occurs mainly during a HIPIMS
pulse, with the growth rate following the HIPIMS phases ignition,
current rise, gas rarefaction, plateau and afterglow. The growth
fluxes of titanium and nitrogen follow slightly different
behaviors with titanium dominating at the beginning of the HIPIMS
pulse and nitrogen at the end of the pulse. This is explained by
the gas rarefaction effect resulting in a dense initial metal
plasma and metal films which are subsequently being nitrified. 
\end{abstract}

\maketitle

\section{Introduction}
Reactive magnetron sputtering is a widely used technique to
deposit many materials such as metals or oxides, carbides, and
nitrides for a variety of applications ranging from hard coatings
for tools, wear resistant coatings in the automotive industry or
bio compatible layers for medical applications. The resulting film
structure is mainly dominated by the dissipated energy per atom
during the growth process, which can be controlled by the ion
bombardment either by varying from a balanced to an unbalanced
magnetron or by additional applying a bias at the substrate level.
This control over film properties would be best for fully ionized
plasmas, however, the required high powers imply a large thermal
load on the targets. A decade ago, this dilemma has been
resolved with the introduction of high power pulse magnetron
sputtering (HIPIMS), where a short high power pulse generates an
almost fully ionized plasma followed by a long off time (duty
cycle typically in the range of percent)\cite{Kouznetsov1999a}.

HIPIMS plasma have also been employed to deposit compounds such as
oxides, nitrides or carbides. These are produced by the addition of
 reactive components such as oxygen, nitrogen or methane 
to the argon sputter gas. The additional reactive gas molecules are
dissociated in the plasma. Together with the sputtered metal from the target,
 a compound is formed on the substrate. These reactive neutrals,
however, may also react with the metal target. This has drastic 
consequences for the process, because the sputter yield at the 
target depends sensitively on the surface composition having usually 
much smaller sputter yield for the poisoned target surface. As a 
consequence, the growth rate decreases drastically and the system 
turns into its poisoned status \cite{Kubart2011}. 
By a proper feedback control of the plasma process this can be
adjusted. The need for feedback control is apparently very much
reduced for HIPIMS processes compared to dcMS processes. This
has been attributed either to the sputter wind emerging from the
target, which results in gas rarefaction and a corresponding lower
flux of reactive neutrals towards the target
\cite{Kubart2011,Rossnagel1998a} or to the intense sputtering of
any compound at the target due to the high voltages of a HIPIMS
process \cite{Wallin2008c}.\\ 

These effects and the interplay between working gas, reactive gas and
target surface might better be understood, when the growth
processes during a single HIPIMS pulse can be temporally resolved.
Film growth in HIPIMS plasmas is an inherent pulsed process with
typical pulse lengths between 50 $\mu$s to 200 $\mu$s. During a HIPIMS 
pulse current and voltages are constantly changing, reaching peak powers 
from $0.5-10.0$~kW/cm$^2$. These IV-characteristics give insight in 
the state of the plasma allowing the discrimination of five distinct
phases \cite{Gudmundsson2012}. 
Particle fluxes per pulse are in the order of 10$^{15}$~cm$^{-2}$s$^{-1}$. 
This means that only 10$^{-4}$ monolayers are deposited per discharge.
Furthermore, the growth rate is not constant during a pulse but changing
with the metalicity of the plasma. It is therefore almost impossible to 
directly measure the varying growth rate during a single pulse by methods 
like ellipsometry or quartz microbalances in the environment of an HIPIMS plasma reactor. 
Recently, we succeeded in devising a method to
measure film growth during a HIPIMS plasma pulse using a rotating
shutter in front of a substrate to spread the temporal development
of the plasma pulse on a lateral scale on a substrate
\cite{Mitschker2012a}. By using that method, the individual growth
processes during titanium sputtering have been resolved
\cite{Mitschker2013}. The individual phases as described by
Gudmundsson et al. \cite{Gudmundsson2012} could be clearly
identified from ignition, current rise, gas rarefaction, plateau
and afterglow. In this paper we develop this method further by
combining it with an X-ray Photoelectron Spectroscopy (XPS) analysis of
the deposition profiles. Thereby, the growth rate can be temporally separated
into a metal part and a compound part.

\section{Experiment}

The rotating shutter experiments are performed in a vacuum vessel
housing a 2 inch magnetron with a titanium target, which is powered
by a Melec HIPIMS power supply. The
substrate consists of a silicon wafer, which is inserted into a 6
inch substrate holder using a load lock system. Argon gas is used
as plasma forming gas at a pressure of 0.25~Pa. A constant HIPIMS
pulse length of 200~$\mu$s  and a duty cycle of 6$\%$ is applied
to the target. The
average power is adjusted to $\langle P \rangle \simeq$ 40~W, while
 the peak power goes up to 0.75~kW. Current and voltage at
the target are monitored with a VI-probe. The VI-probe consists
of a LEM LA305-S current transducer with a response time 
$t_r<1$~$\mu$s and a d$I/$d$t>$100~A/$\mu$s and a LEM CV3-1500 
voltage transducer with $t_r=0.4$~$\mu$s and d$V/$d$t=$900~V/$\mu$s. 
The measurements are taken directly at the output of the HIPIMS power supply.\\

Nitrogen is used as reactive gas for all experiments. A control 
of the hysteresis effect is performed by an active control system 
for the nitrogen flux. This system monitors the intensity of the titanium
line at 307~nm (3d$^2$4p $\to$ 3d$^2$4s). The intensity 
of this line is damped with increasing nitrogen concentration. It is
therefore used as actuating variable to regulate the reactive gas influx.
The optical emission of the HIPIMS-plasma is monitored by a Hamamatsu RCA 78-
09 IP28 photomultiplier. The photomultiplier is directed towards the center 
between the magnetron and the substrate. An optical filter with a center 
wavelength of 307$\pm 2$~nm and a full width half maximum of 10$\pm 2$~nm 
is placed in front of the photomultiplier in order to isolate the titanium
line of interest. The signal of the photomultipler is processed by an 
analog integrator and an amplifier followed by an analog to digital converter. 
This signal is the input parameter for a software PID-controller which then regulates 
the reactive gas influx via a fast-acting Pfeiffer RME 005 A valve. 
In all experiments, the emission intensity was kept at $15\%$ of 
the maximum emission intensity for the whole deposition time.\\

In front of the substrate, a rotating shutter is placed to measure
the growth flux on the substrate. According to that concept a 
200~$\mu$m slit is rotated in front of
the substrate synchronous with the HIPIMS pulses. Thereby, the
growth flux is laterally distributed over the substrate. By
measuring the resulting deposition profile with profilometry and
with x-ray photoelectron spectroscopy, the temporal variation of
the titanium and nitrogen growth flux per pulse is deduced. For
details see F. Mitschker et al. \cite{Mitschker2013}.

In this study the deposition time was six hours.
After deposition, the substrates are extracted and
the film deposition profiles on the wafers are measured using
profilometry. Profilometry is most exact, if a step edge between
substrate and film can be measured. Such a step edge is generated
by placing at first a vacuum tape on the substrate. After
deposition, this vacuum tape is removed leaving a sharp
edge behind, which is evaluated using profilometry with an 
accuracy of $\pm$ 10 nm.\\

By using an experiment with a stopped shutter, the minimum spread
of the deposition profile is measured, which can be converted into
a minimum temporal resolution of 25 $\mu$s for the shutter
rotating at 30 Hz. Details of this calibration can be found
elsewhere \cite{Mitschker2013}.

The composition of the deposition profiles is ex-situ measured by XPS. 
These measurements are performed in 
a Versaprobe Spectrometer from Physical Electronics, using 
monochromatic Al K$\alpha$ radiation at 1486.6~eV and a pass energy of 
23.5~eV, which results in a full-width at half-maximum (FWHM) 
of 0.58~eV for the Ag3d$_{5/2}$ core level of a clean silver sample. 
The measurement spot has a diameter of 200~$\mu$m. The TiN samples 
are introduced to the measurement chamber without any prior cleaning. 
Due to the ex-situ measurement the samples may react with ambient
air. We assume that any Ti atom not saturated by N may react with O$_2$
to form TiO-groups. The surface of the sample is not 
neutralised. It is expected that the samples show good conductivity.
Further, the relative position of photoelectron lines is independent 
of the sample composition. The take-off angle between sample surface 
and analyzer level is 45$^\circ$.

Measurements are done parallel to the deposition profile, where in a 
first step a survey spectrum was taken every 1~mm. Markings on the silicon 
wafer serve as a reference position for XPS and profilometry. The N1s 
photoelectron line is used to define the measurement spots for a more 
detailed investigation. In regions, in which N1s is observed, the distance 
between two measurement spots is chosen as 200~$\mu$m, while it is 
1000~$\mu$m for the outer parts where no nitrogen is found. Using the 
MultiPak\texttrademark{} software, the ratios of N1s, O1s and Ti2p are
 determined by weighting the data with with their individual sensitivity 
factors. These composition profiles can be used to decompose the deposition 
profile, obtained by profilometry, into its individual components.
\section{Results and Discussion}
\subsection{Deposition profile for titanium and nitrogen}
\label{sec:results_and_discussion}
\begin{figure}
  \centering
  \includegraphics[width=10cm]{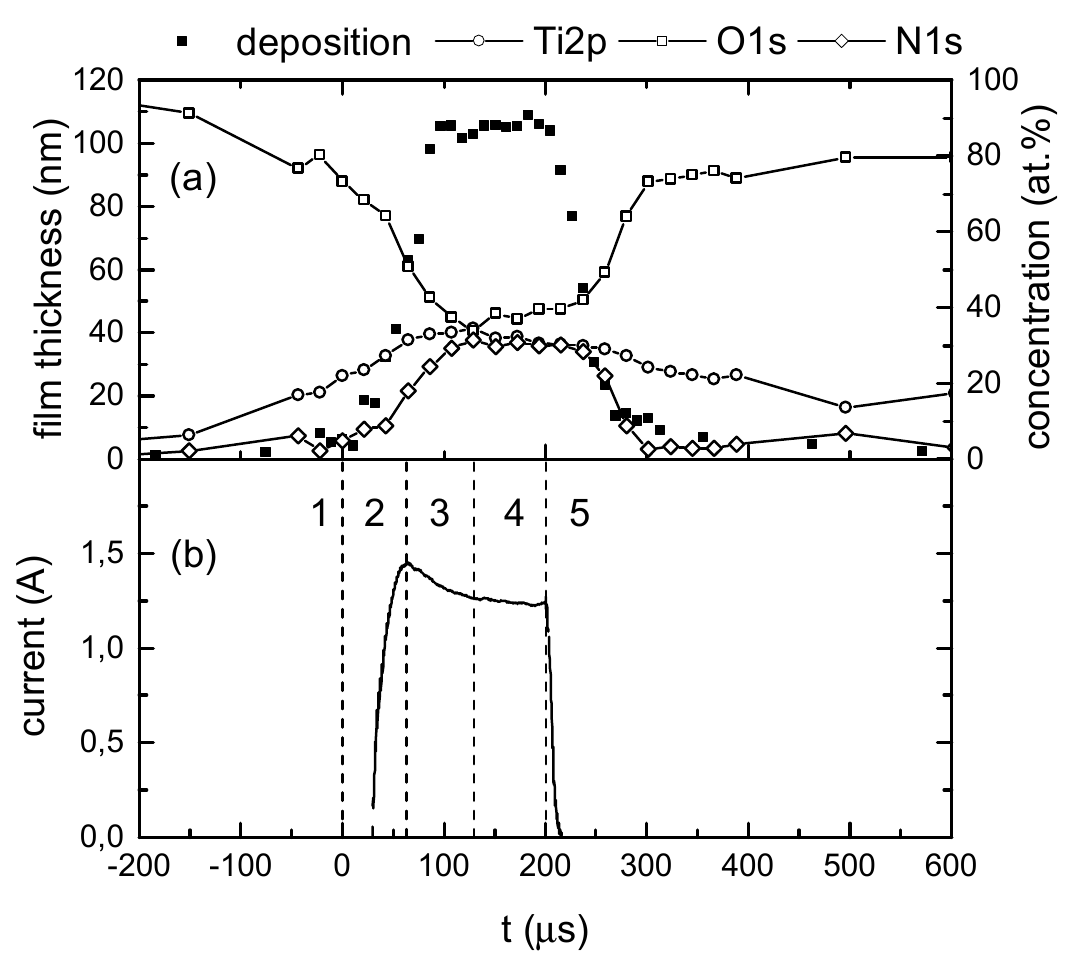}
  \caption{(a) Deposition profile (solid symbols),
  concentration of titanium, oxygen, and nitrogen (open symbols);
  (b) Smoothed current during a HIPIMS pulse of 200~$\mu$s
  with a target voltage of $V_{target}=467$~V at an
  average power of $\langle P\rangle$=40~W
  at a pressure of $p$ = 0.25 Pa. The numbers indicate
  different phases of the growth process. For details see text.
  }
 \label{figure2}
\end{figure}

The deposition profile has been measured for 200 $\mu$s pulse with
6$\%$ duty cycle at a
pressure of 0.25~Pa and $\langle P \rangle \simeq$ 40~W, as 
shown in Fig.~\ref{figure2}(a). The start and the end
of the pulse is indicated with dashed lines. The profiles are synchronised in time
by regarding the end of the current pulse. 
Some deposition might be indicated before plasma ignition, although it is 
only caused by the finite width of the deposition profile \cite{Mitschker2013}.
The base pressure in the chamber was $4\times 10^{-5}$~Pa. In order to control the 
hysteresis effect, the intensity of the Ti-line at 307~nm was kept on $15\%$ of the maximum
 emission intensity for the whole deposition time via the described control mechanism.
The deposition took six hours to reach a maximum thickness of about 110~nm.\\

The evolution of the smoothed current is presented in Fig. \ref{figure2}(b) with the
individual HIPIMS phases indicated as 1 (ignition), 2 (current
rise)\cite{Lundin2011}, 3 (gas rarefaction) \cite{Hala2010,Rossnagel1998a}, 
4 (plateau) \cite{Anders2007}, and 5 (afterglow) \cite{Poolcharuansin2010}. 
The deposition rate follows these phases in a characteristic manner.
At first, the plasma starts (phase 1), the current
rises steeply (phase 2) and finally goes through a maximum and
declines again (phase 3). At the same time, the
growth rate increases only slowly in phase 2. A strong increase of
the growth rate is observed in phase 3 followed by a
saturation at a constant value (phase 4). After the end
of the pulse, the growth rate decreases sharply with a time
constant to the order of 100 $\mu$s followed by a tail with a 
slower decay constant (phase 5). For details see \cite{Mitschker2013}.

This temporal evolution of the growth rate can now be separated
into the contribution of titanium and nitrogen in the film by
measuring along the deposition profile with XPS. Fig.~\ref{figure2}a 
shows the corresponding concentrations in at.$\%$ of titanium,
nitrogen, and oxygen. The surface oxygen is assumed to result from
post-oxidation in the ambient atmosphere due to the ex-situ XPS
analysis. The oxygen concentration is the highest in regions with
an excess of titanium in comparison to nitrogen. Apparently, non
stoichiometric TiN parts of the deposition profile are oxidized
ex-situ after contact with ambient air. Consequently, we discuss
in the following only the ratio between titanium and nitrogen as
it is directly created in-situ during the HIPIMS pulse.\\

One observation is that the titanium and the nitrogen
profile do not exhibit the same shape. At the beginning and at the
end of the deposition profile, the surface is titanium-rich with a
very low concentration of nitrogen. Such a behavior may be
consistent with the gas dynamics in a HIPIMS pulse:

\begin{itemize}
\item {\it begin of the HIPIMS pulse}: At the beginning of the
plasma in phase 1 and 2, a strong gas rarefaction sets in, because
the sputtering wind expels the neutral reactive gas in front of
the target \cite{Rossnagel1998a}. Consequently, the flux of
titanium is high and that of nitrogen low. Later in the pulse, the
fluxes of titanium and nitrogen come to an equilibrium after the
initial current peak goes into the plateau phase 4 and the
nitrogen uptake of the surface can take place.

\item {\it end of the HIPIMS pulse}: At the end of the pulse, the 
nitrogen concentration drops much sharper 
than the titanium concentration. One may speculate that the incorporation 
of nitrogen into the layers is dominated by incident nitrogen ions rather 
than neutrals, because only ions disappear on a short time scale of the 
order of 100 $\mu$s. A low sticking coefficient for incident nitrogen atoms is 
a reasonable assumption.
\end{itemize}

It is important to note that the deposited atoms before and after
the HIPIMS pulse constitutes only to a small fraction to the total
amount of deposited material. Nevertheless, it indicates that the
HIPIMS process is a sequential process where a titanium rich
layer is followed by a TiN layer. If one regards the total fluxes
per pulse, lasting 200~$\mu$s and regarding a growth flux of the
order of 10$^{15}$~cm$^{-2}$s$^{-1}$, a fluence per pulse of only
$\sim 10^{11}$~cm$^{-2}$ is accumulated during each pulse. This is only
a small fraction of a monolayer and any structural consequences
from the layer-by-layer growth mode for the deposited TiN films is
not expected. However, the electronic structure of the surface might 
still be affected by even a small contribution of the sequential growth
mode, since the work function might be altered even by small
concentrations.
\subsection{Dependence on pressure}
\begin{figure}
  \centering
  \includegraphics[width=10cm]{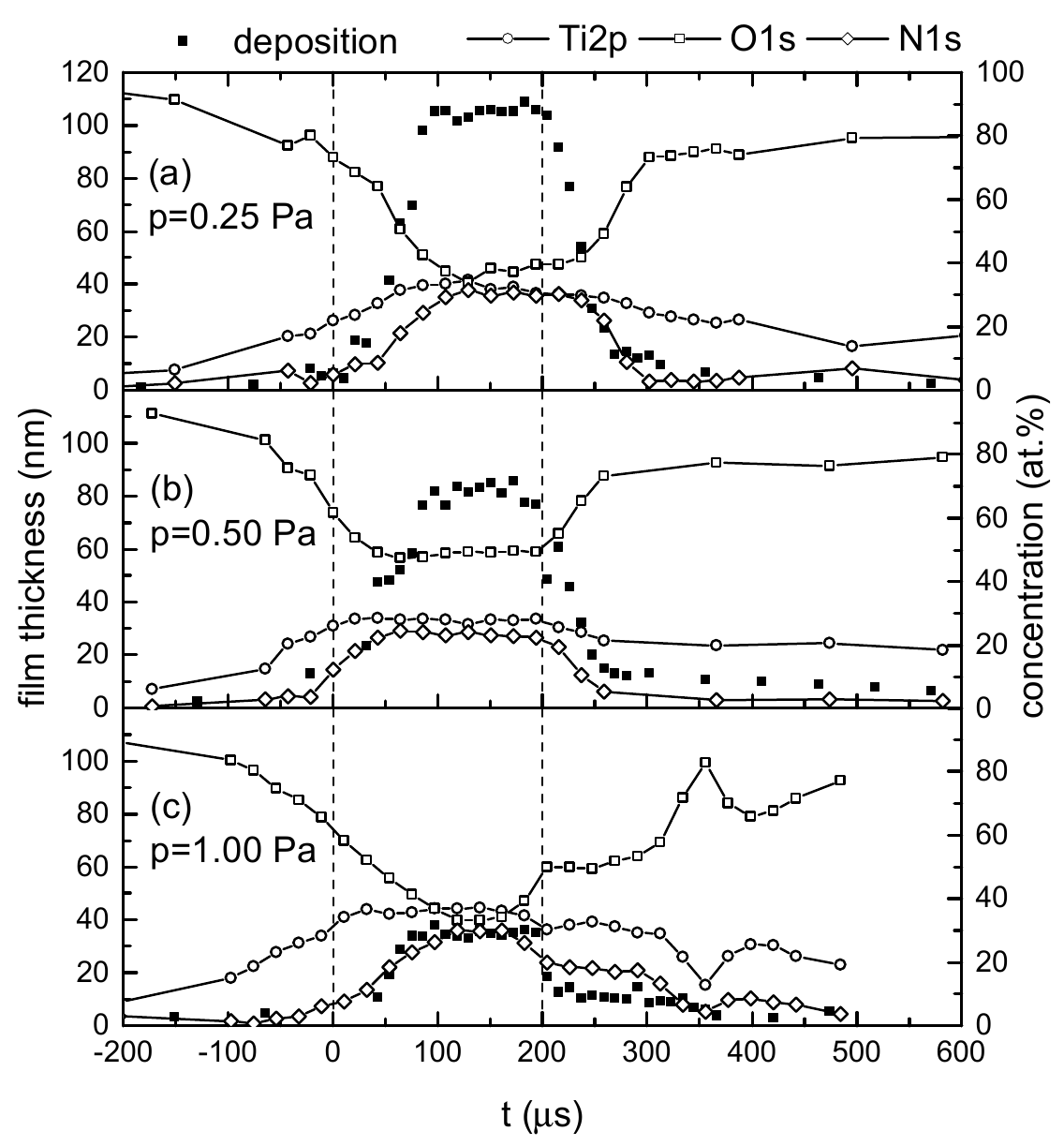}
  \caption{Deposition profile (solid symbols), concentration of titanium,
  oxygen, and nitrogen (open symbols)
  at a pressure of $p$ = 0.25 Pa (a), $p$ = 0.5 Pa (b) and $p$ = 1 Pa (c) 
and a constant power of $\langle P \rangle \simeq40$~W. The vertical dashed 
lines indicate the start and the end of the HIPIMS pulse.
  }
 \label{figure3}
\end{figure}

The deposition profiles are further analysed for pressures 
of 0.25, 0.5 and 1~Pa, as shown in Fig.~\ref{figure3}. All experiments 
are performed at an average power of $\langle P \rangle \simeq40$~W, a 
pulse width of 200~$\mu s$ and a duty cycle of 6$\%$. This results in 
different target voltages for different pressures. Consequently, all 
experiments correspond to different HIPIMS plasma pulses. The deposition 
time for each sample was six hours.
\\

\begin{figure}
  \centering
  \includegraphics[width=10cm]{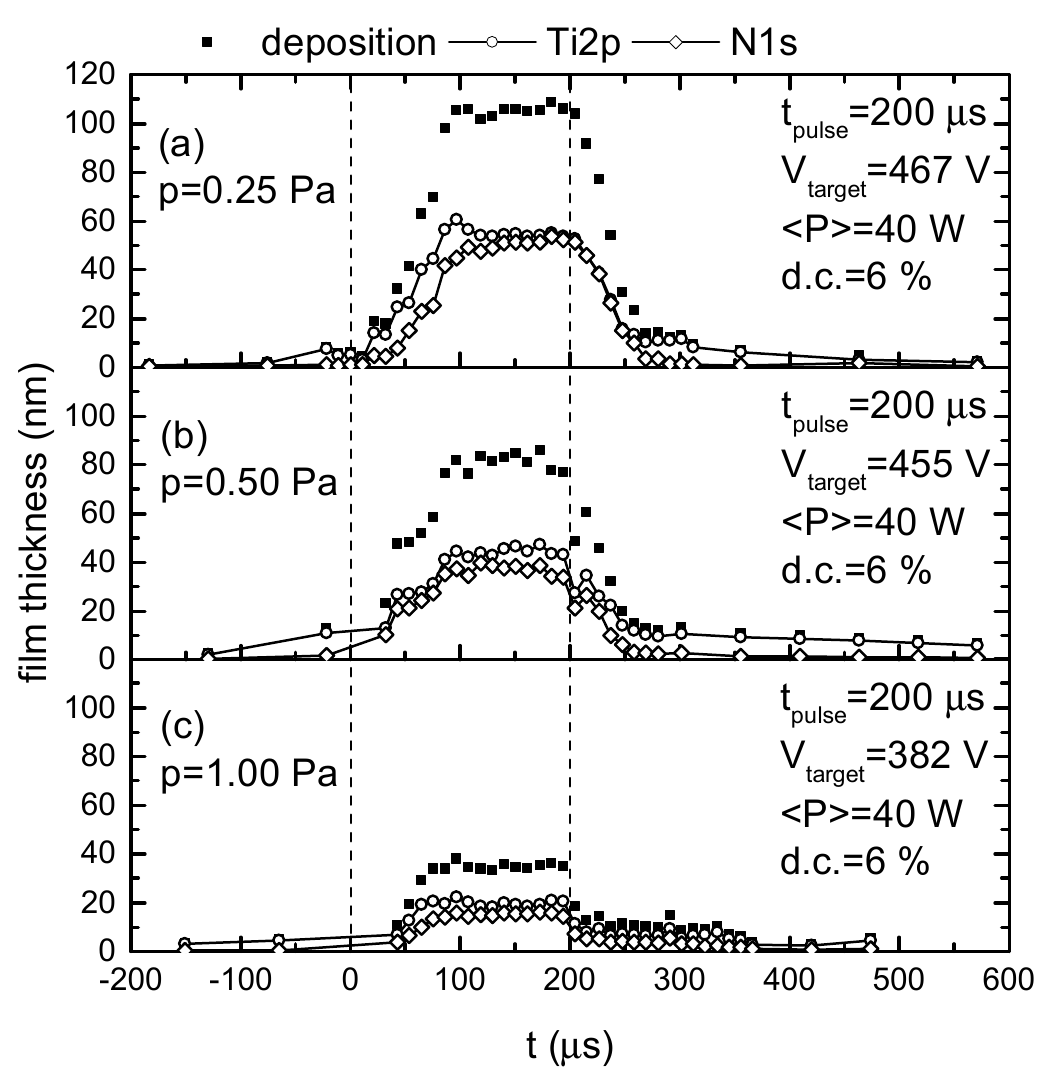}
  \caption{Deposition profile (solid symbols) and disentangled deposition profiles 
for titanium (open circles) and nitrogen (open diamonds) at a pressure of $p$ = 0.25 Pa (a)
, $p$ = 0.5 Pa (b) and $p$ = 1 Pa (c) and a constant power of $\langle P \rangle \simeq40$~W.
 The vertical dashed lines indicate the start and the end of the HIPIMS pulse.
  }
 \label{figure4}
\end{figure}

With the knowledge of the concentrations of the different species 
(Fig.~\ref{figure3}), it is possible to disentangle the total deposition 
profile in a titanium and nitrogen profile. (Fig.~\ref{figure4}). All three 
measurements are consistent with the characteristics described in 
Sec.~\ref{sec:results_and_discussion}. At the beginning and the end of the 
growth pulse, the profiles are titanium-rich for all three pressures. The biggest 
difference is in the pressure dependence of the growth rate. With a pressure 
increase from 0.25~Pa to 1.00~Pa the growth rate diminishes almost by a factor of 
three. This indicates the transition from a ballistic to a diffusive transport 
of the species from target to substrate \cite{Mitschker2013}.\\

\section{Conclusion}
The dynamics of the growth rate and the composition of the resulting 
films are consistent with the current understanding of the physics of a HIPIMS 
pulse as being expressed by the five phases from ignition to afterglow. 
It is observed, that the deposition profile is titanium rich with a low
nitrogen content at the beginning and the end of the pulse. This behavior
seems to be coherent with the gas dynamics of a HIPIMS pulse. The temporal 
evolution of the growth rate is dominated by the dynamic change of the transport 
properties between collisional to ballistic transport whereas the composition depends 
on the gas depletion and the metalicity of the plasma.\\
 Summarizing, one can state that 
the rotating shutter method is an excellent tool to resolve the time dependence of 
growth processes in pulsed plasmas. This will help to elucidate the nature of the
HIPIMS growth process in the future.
\section*{Acknowledgements}

The authors would like to thank Norbert Grabkowski for his
technical support. This project is supported by the DFG (German
Science Foundation) within the framework of the Coordinated
Research Center SFB-TR 87 and the Research Department `Plasmas
with Complex Interactions' at Ruhr-University Bochum. Finally, we
would like to thank Ante Hecimovic for helpful discussions.


\section*{References}

\end{document}